\documentstyle[multicol,aps,prb,epsf]{revtex}

\begin{document}

\title{Ground state properties of fluxlines in a disordered environment}

\author{Heiko Rieger}
\address{HLRZ, Forschungszentrum J\"ulich, 52425 J\"ulich, Germany}

\date{June 30, 1997}

\maketitle

\begin{abstract}
  A new numerical method to calculate exact ground states of
  multi-fluxline systems with quenched disorder is presented, which is
  based on the minimum cost flow algorithm from combinatorial
  optimization. We discuss several models that can be studied with
  this method including their specific implementations, physically
  relevant observables and results: 1) the $N$-line model with $N$
  fluxlines (or directed polymers) in a $d$-dimensional environment
  with point and/or columnar disorder and hard or soft core repulsion;
  2) the vortex glass model for a disordered superconductor in the
  strong screening limit and 3) the Sine-Gordon model with random pase
  shifts in the strong coupling limit.
\end{abstract}

\pacs{PACS numbers: 02.60.Pn, 36.20-r, 74.60.Ge, 64.60.Cn}

\newcommand{\bc}{\begin{center}}
\newcommand{\ec}{\end{center}}
\newcommand{\be}{\begin{equation}}
\newcommand{\ee}{\end{equation}}
\newcommand{\beqn}{\begin{eqnarray}}
\newcommand{\eeqn}{\end{eqnarray}}

\begin{multicols}{2}
\narrowtext
\parskip=0cm

Dirty type II superconductors in a magnetic field are the most
intensively studied representatives of elastic manifolds in a
disordered environment \cite{vortexrev}. Their paradigmatic
description consists in an ensemble of magnetic fluxlines (or
vortexlines) interacting strongly with point and/or columnar defects
and among themselves. This complicated multi-line situation is usually
reduced to the study of a single line, a directed polymer in a random
medium \cite{dp_review}, a problem that possesses deep connection also
to nonequilibrium fluctuations of moving interfaces \cite{kpz}.

Here we are going to present a new numerical method (in the spirit of
other recent applications of combinatorial optimization tools in the
physics of disordered systems\cite{review}) by which the investigation
of the full multi-line situation becomes feasible. It will enable us
to determine exact ground sates (i.e.\ minimum energy configurations)
of theses systems in polynomial time. Since the low temperature
physics of fluxlines in a random environment is dominated by disorder
effects these ground state calculations will enable us to make various
statements about possibly glassy features, for instance the roughness
of multi-line systems, the stiffness of vortex or gauge glass models
and the displacement-displacement correlations in random phase models.

To introduce the notation and to set the stage of the theoretical
models we consider we start with a simple but non-trivial (and hence
heavily discussed \cite{dp_review}) example: the so called $1$-line
problem, which consists in determining the minimum energy
configuration of a single (magnetic) fluxline or a directed polymer
(for a 111-lattice) in a disordered environment. The lattice version
of this model is given by the Hamiltonian (or energy function)
\be
H({\bf x})=\sum_{(ij)} e_{ij}\cdot x_{ij}\;,
\label{hamilflux}
\ee
where $\sum_{(ij)}$ is a sum over all {\it bonds} $(ij)$ joining site
$i$ and $j$ of a $d$-dimensional lattice, e.g.\ a rectangular
($L^{d-1}\times H$) lattice, with periodic boundary conditions (b.c.)
in $d-1$ space direction and free b.c.\ in one direction. The bond
energies $e_{ij}\ge0$ are quenched random variables that indicate how
much energy it costs to put a segment of fluxline on a specific bond
$(ij)$.  The fluxline configuration ${\bf x}$ ($x_{ij}\ge0$), also
called a {\it flow}, is given by specifying $x_{ij}=1$ for each bond
$i$, which is occupied by the fluxline and $x_{ij}=0$ otherwise. For
the configuration to form {\it lines} on each site of the lattice all
incoming flow should balance the outgoing flow, i.e.\ the flow is
divergence free
\be
\nabla\cdot{\bf x}=0\;,
\label{divfree}
\ee
where $\nabla\cdot$ denotes the lattice divergence. Obviously the
fluxline has to enter, and to leave, the system somewhere.  We attach
all sites of one free boundary to an extra site (via energetically
neutral arcs, $e=0$), which we call the source $s$, and the other side
to another extra site, the target, $t$ as indicated in fig.\ 1a. Now
one can push one line through the system by inferring that $s$ has a
source strength of $+1$ and that $t$ has a sink strength of $-1$,
i.e.\
\be
(\nabla\cdot{\bf x})_s=+N\quad{\rm and}\quad
(\nabla\cdot{\bf x})_t=-N\;,
\label{stdiv}
\ee
with $N=1$. Thus, the $1$-line problem consists in minimizing the
energy (\ref{hamilflux}) by finding a flow ${\bf x}$ in the network
(the lattice plus the two extra sites $s$ and $t$) fulfilling the
constraints (\ref{divfree}) and (\ref{stdiv}).

The solution of this problem is equivalent to finding the shortest
path from $s$ to $t$, where distances between two lattice sites are
identified with the energies $e_{ij}$, which can either be done with
Dijkstra's algorithm from combinatorial optimization\cite{flows} or by
equivalent methods better known to physicists: the transfer matrix
method \cite{transfer}.

Since this $1$-line problem has been extensively
studied\cite{dp_review} we directly proceed to its full generalization
to $N$ fluxlines, which has, to our knowledge, never been treated in
the literature before. The reason is simple: Whereas two lines ($N=2$)
are still tractable \cite{tang}, the transfer matrix method fails to
work efficiently for an increasing number of lines since its
complexity grows exponentially with $N$. Since it is the dense limit
$N=\rho L^{d-1}$ with $\rho$ of order one which is expected to contain
new physics an algorithm that solves this problem in polynomial time
as the one we are now going to present, is highly desirable.

The $N$-line problem again consists in minimizing (\ref{hamilflux}) in
such a way that (\ref{divfree}) and (\ref{stdiv}) are fulfilled, now
with an arbitrary value for $N$.  Physically one has to take into
account a repulsive interaction between the fluxlines, for instance a
hard core repulsion, which can be modeled by inferring that
$x_{ij}\in\{0,1\}$, i.e.\ that only a segment of one single fluxline
can pass through an arc \cite{remark}. It is also possible to apply
our method to a situation with soft core repulsion, which we discuss
below. The problem is now formulated in such a way that it is
identical to minimum cost flow problem in combinatorial optimization
\cite{flows}.

Since the 1-line problem can be solved by finding a shortest path the
intuitive idea to solve the $N$-line problem would be to search
successively $N$ shortest path, i.e.\ by adding one fluxline after the
other to the system.  However, adding a fluxline to an existing
fluxline configuration might necessitate redirecting one or more
fluxlines, as indicated in fig.\ 1.  This at first sight formidable
task is elegantly solved by the so called successive shortest path
algorithm for minimum cost flow problems \cite{flows}.

\begin{figure}
\epsfxsize=\columnwidth\epsfbox{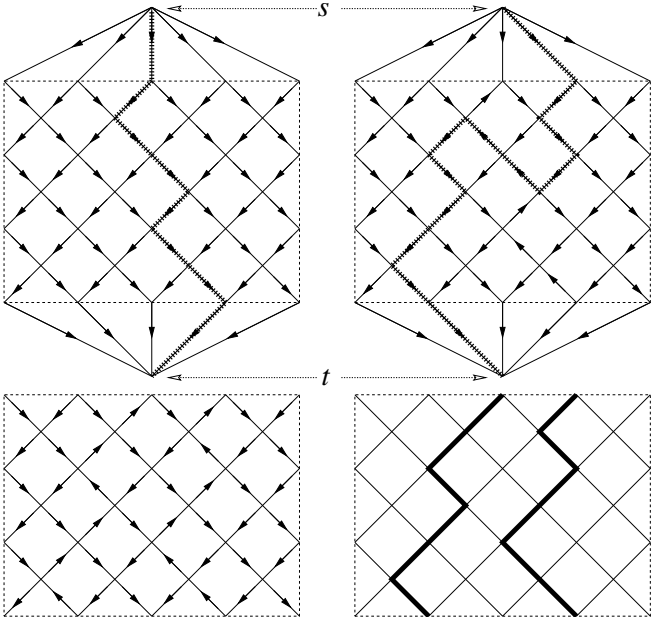}
\caption{
\label{fig1}
Sketch of the residual network with shortest path before putting in th
e first fluxline (top left) and after the update for one fluxline (top
right). Bottom: Residual network after second iteration (left) and 
the actual optimal fluxline configuration (right). Note that this case
is a non-trivial case when the first fluxline has to be deformed in
order to achieve the optimal 2-line configuration.}
\end{figure}

The first key ingredience is that one does not work with the original
network but with the residual network corresponding to the actual
fluxline configuration, which contains also the information about
possibilities to send flow backwards (now with energy $-e_{ij}$ since
one wins energy by reducing $x_{ij}$), i.e.\ to modify the actual flow.
Suppose that we put one fluxline along a shortest path $P(s,t)$ from
$s$ to $t$, which means that we set $x_{ij}=1$ for all arcs on the
path $P(s,t)$. Then the residual network is obtained by reversing all
arcs and inverting all energies along this path, indicating that here
we cannot put any further flow in the forward direction (since we
assume hard-core interaction, i.e.\ $x_{ij}\le1$), but can send flow
backwards by reducing $x_{ij}$ on the forward arcs by one unit. This
procedure is sketched in figure 1.

The second key ingredience is the introduction of a so called
potential ${\bf\varphi}$ that fulfills the relation
\be 
\varphi(j)\le \varphi(i)+e_{ij} 
\label{potential}
\ee 
for all arcs $(ij)$ in the residual network, indicating how much
energy $\varphi(j)$ it would {\it at least} take to send one unit of flow
from $s$ to site $j$, IF it would cost an energy $\varphi(i)$ to send it
to site $i$. With the help of these potentials one defines the reduced
costs
\be
c_{ij}^{\bf\varphi}=e_{ij}+\varphi(i)-\varphi(j)\ge0\;.
\label{redcost}
\ee
The last inequality, which follows from the properties of the
potential ${\bf\varphi}$ (\ref{potential}) actually ensures that there
is no loop ${\cal L}$ in the current residual network (corresponding
to a flow {\bf x}) with negative total energy, since
$\sum_{(ij)\in{\cal L}}e_{ij} = \sum_{(ij)\in{\cal
    L}}c_{ij}^{\bf\varphi}$, implying that the flow ${\bf x}$ is
optimal\cite{flows}. 

It is important to note that the inequality (\ref{potential}) is
reminiscent of a condition for shortest path distances $d(i)$ from $s$
to all sites $i$ with respect to the energies $e_{ij}$: they have to
fulfill $d(j)\le d(i)+e_{ij}$. Thus, one uses these distances $d$ to
construct the potential ${\bf\varphi}$ when putting one fluxline after
the other into the network:

We start with the empty network (zero fluxlines) ${\bf x}^0=0$, which
is certainly an optimal flow for $N=0$, and set ${\bf\varphi}=0$,
$c_{ij}^{\bf\varphi}=e_{ij}$. Next, let us suppose that we have an
optimal $N-1$-line configuration corresponding to the flow ${\bf
  x}^{N-1}$.  The current potential is ${\bf\varphi}^{N-1}$, the
reduced costs are
$c_{ij}^{N-1}=e_{ij}+\varphi^{N-1}(i)-\varphi^{N-1}(j)$ and we
consider the residual network $G_c^{N-1}$ corresponding to the flow
${\bf x}^{N-1}$ with the reduced costs $c_{ij}^{N-1}\ge0$. The
iteration leading to an optimal $N$-line configuration $x_{ij}^{N}$ is

\itemsep=0cm
\begin{itemize}
\item[1.] Determine a shortest path $P(s,t)$ with respect to the
          reduced costs $c_{ij}^{N-1}$from $s$ to $t$ in the residual
          network $G_c^{N-1}$.
\item[2.] For all site on $P(s,t)$ let $d(i)$ be the shortest path
          distance from $s$ to $i$. For these update the potentials: 
          $\varphi^{N}(i)=\varphi^{N-1}(i)+d(i)-d(t)$.
\item[3.] To obtain $x_{ij}^{N}$ increase (decrease) by one unit 
          the flow variables $x_{ij}^{N-1}$ on all forward (backward) 
          arcs $(ij)$ on the shortest path $P(s,t)$.
\end{itemize}

Note that due to the the fact that the numbers $d(i)$ are shortest
distances one has again $c_{ij}^{N}\ge0$, i.e.\ the flow ${\bf
x}^{N}$ is indeed optimal. The complexity of this iteration is the
same as that of Dijkstra's algorithm for finding shortest paths in a
network, which is ${\cal O}(M^2)$ in the worst case ($M$ is the number
of nodes in the network). We find, however, for the cases we consider
($d$-dimensional lattices) it roughly scales linear in $M=L^d$. Thus,
for $N$ fluxlines the complexity of this algorithm is ${\cal
O}(NL^d)$.
\begin{figure}
\epsfxsize=\columnwidth\epsfbox{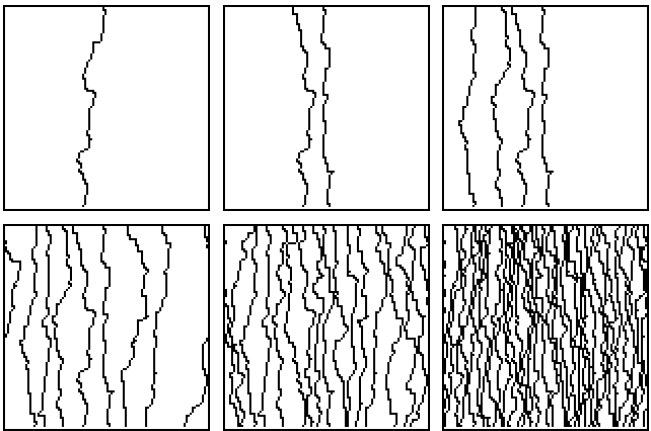}
\caption{
\label{fig2}
Optimal fluxline configuration for one particular 2d sample of linear
size $L=100$. The number of fluxlines, $N$, is 1,2,4 (top), 8,16,32 (bottom).}
\end{figure}
\begin{figure}
\epsfxsize=\columnwidth\epsfbox{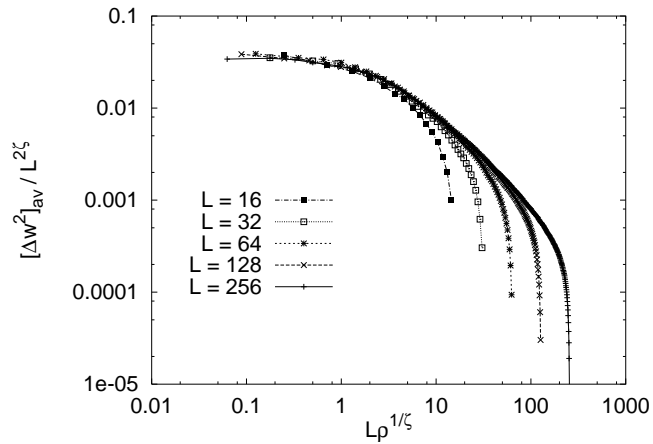}
\caption{
\label{fig3}
Scaling plot for the roughness (per line) in the two-dimensional
multi-fluxline system, cf.(\ref{roughscale}). The data are averaged
over 1000 different disorder configurations.
}
\end{figure}
In fig.\ 2 we show a number of optimal fluxline configurations with
varying line density obtained with our algorithm for the $N$-line
problem on a $L\times L$ square lattice. We would like to emphasize
the algorithm works for fluxlines in arbitrary dimension, even for
general graphs. One observes that the roughness $[\Delta w]_{\rm av}$
of the fluxlines decreases systematically with increasing line density
from the single line limit ($N=1$) $[\Delta w]_{\rm av}\sim L^\zeta$
with the roughness exponent $\zeta=2/3$ to the dense limit ($N=L$ in
$d=2$) with no roughness at all, $[\Delta w]_{\rm av}=0$. For a finite
fluxline density $\rho=N/L$ each fluxline is essentially free up to a
length $\xi_\parallel=\xi_\perp^{1/\zeta}$ with $\xi_\perp=1/\rho$.
Thus one expects for $d=2$ the finite size scaling form
\be
[\Delta w]_{\rm av}\sim L^\zeta\tilde{w}(\,H\rho^\nu\,)\quad
{\rm with}\quad\nu=1/\zeta
\;,
\label{roughscale}
\ee
where $H$ is the height and $\tilde{w}$ is a scaling function with
$\tilde{w}(x)\to const.$ for $x\to0$. In fig.\ 3 we show a
corresponding scaling plot for the data obtained with our algorithm
for $d=2$ (and $H=L$) \cite{viljo}.

Before we proceed we would like to point out that the Hamiltonian
(\ref{hamilflux}) is general enough to describe various physically
interesting situations. By an appropriate definition of the energies
$e_{ij}$ one can easily model columnar defects\cite{columnar},
disorder induced melting of a fluxline lattice (e.g.\ the Abrikosov
lattice) and depinning transitions.


Next we consider soft core repulsion, which can be modeled by
allowing a multiple occupancy of a bond ($x_{ij}=0,1,2,\ldots$) but
punish high fluxline densities with an energy $\tilde{e}_{ij}(x_{ij})$
increasing faster than linear with the number of flux units $x_{ij}$
on the bond $(ij)$.  Thus the $N$-line problem with soft repulsion
consists in minimizing
\be
\tilde{H}({\bf x})=\sum_{(ij)} \tilde{e}_{ij}(x_{ij})\;,
\label{convexflow}
\ee
under the constraints (\ref{divfree}) and (\ref{stdiv}). The local
energy functions $\tilde{e}_{ij}$ can be chosen arbitrarily for each
bond $(ij)$, however, they have to be {\it convex} as for instance
$\tilde{e}_{ij}(x_{ij})=k_{ij}\cdot x_{ij}^n$ with $n\ge1$ arbitrary.
The energies $e_{ij}$ have now to be replaced by the quantity
$\tilde{e}_{ij}(x_{ij}+1)-\tilde{e}_{ij}(x_{ij})$, which is the energy
needed to increase the flow $x_{ij}$ on arc $(ij)$ by one unit. Since
it depends on the current flow ${\bf x}$ the convexity of
$\tilde{e}_{ij}$ is needed to ensure that the reduced costs fulfill
the inequality $c_{ij}^{N}\ge0$ (\ref{redcost}) also after the flow
modification. Whereas with hard core repulsion it was only possible to
put $N=L^d$ fluxline into the system, the fluxline density can now
arbitrarily high and an interplay between the repulsion and the
disorder effects lead to a much richer phenomenology \cite{viljo}.

Up to now we considered situations in which the fluxlines are put into
the system via an explicit external source. We now present a model in
which fluxlines are generated {\it inside} the system: namely the
vortex representation of the gauge glass model \cite{gauge} with
strong screening, which is given by the Hamiltonian \cite{WY}
\be
H_V=\sum_{(ij)} (x_{ij} - b_{ij})^2 \,.
\label{vortex}
\ee
Here $\sum_{(ij)}$ is a sum over all bonds of a simple cubic lattice
($d=3$) with periodic b.c.\ in all directions and we do {\it not} have
external source nodes subjected to condition (\ref{stdiv}). The
$x_{ij}$ are the integer flow variables that have to fulfill the
divergence free condition (\ref{divfree}), and the $b_{ij}$ are
quenched random variables that are {\it real} numbers. They can be
arbitrary, however in the gauge glass they fulfill a divergence free
condition $\nabla\cdot{\bf b}=0$ since they represent a magnetic field
derived from a quenched random vector potential (${\bf
  b}=\nabla\times{\bf A}$). Without the constraint (\ref{divfree}) the
optimal solution would simply be given by choosing $x_{ij}$ to be the
closest integer to $b_{ij}$. This solution fulfills
$c_{ij}^{\bf\varphi}\ge0$ with ${\bf\varphi}=0$, where the costs are
chosen as for the convex flow problem (\ref{convexflow}). Since it
violates the constraints (\ref{divfree}) one has either excess or
deficit flow entering or leaving individual sites, which one has to
remove. Instead of sending flow from one particular source node
$s$ to a target $t$ as in the fluxline problem, one now sends flow
from excess to deficit sites along shortest paths using the iteration
described above. In this way one successively removes the violations
of constraint (\ref{divfree}) by keeping the reduced cost optimality
$c_{ij}^{\bf\varphi}\ge0$ all the time, which guarantees the
optimality of the flow at the end of the iteration.

The physically most interesting question in the context of the model
Hamiltonian (\ref{vortex}) concerns the existence of a superconducting
glass phase. This can be studied via domain wall renormalization group
methods \cite{DWRG}, by which one determines the scaling behavior of
low lying excitation $\Delta E$ on the length scale $L$. Such an
excitation in the vortex representation (\ref{vortex}) is a loop
(closed fluxline) with an area proportional to $L^2$, which can be
realized by an extra fluxline (on the background of the true ground
state) winding once around the 3d torus in one direction (note that we
have periodic b.c.\ in all directions). Details of this procedure will
be published elsewhere, here we give only the result:
$\Delta E \sim L^{\theta}$ with $\theta = -0.95 \pm 0.03$
%
%
%
From this we can draw two conclusions: a) since it is clearly negative
there is no superconducting (vortex) glass phase at non-vanishing
temperature, and b) the thermal correlation length diverges only with
$T\to0$ as $\xi_{\rm th}\sim T^{-\nu}$ with an exponent
$\nu=1/\vert\theta\vert\sim1.05\pm0.03$, which is in agreement with
Refs.\ \cite{WY}.

Finally, as our last application of our algorithm we discuss the
Sine-Gordon model with random phase shifts, which in two dimensions is
a model for a fluxline array \cite{batrouni}
\be
H_{RP}=\sum_{(ij)} (u_i-u_j)^m
-\lambda\sum_i\cos\,\bigl(2\pi(u_i-\beta_i)\bigr)\;,
\label{rf}
\ee
where $u_i$ are (real) displacements in a fluxline array, 
$\beta_i\in[0,1[$ (quenched) random phase shifts, $(ij)$
nearest neighbor pairs of a $d$-dimensional lattice, $m\ge1$ is some
number (usually it is $m=2$, by which it the first term in
(\ref{rf}) becomes an elastic energy) and $\lambda$ a coupling
parameter. In the limit of strong coupling $\lambda\to\infty$ one
enforces $u_i=\beta_i+n_i$, where $n_i$ is an integer, and the
Hamiltonian reads in these variables $H_{RP}'=\sum_{(ij)}
(n_i-n_j-\theta_{ij})^m$,
%
%
where $\theta_{ij}=\theta_i-\theta_j$. In $d=2$ this model is
equivalent to a model for a crystalline surface on a random substrate
\cite{sos}, for which the variables $n_i$ are interpreted as integer
height variables and $\theta_i$ as substrate heights between $0$ and
$1$.

By introducing the variables $x_{ij}=n_i-n_j$ one observes
that the form of the Hamiltonian is a special case of the
convex cost functions considered above (\ref{convexflow}) for which
the ground state can again computed with the algorithm we have
presented. However, now one has to consider a simply connected
topology for the underlying network (e.g.\ {\it not} the torus) since
we have to reconstruct the variables $n_i$ from their differences
$x_{ij}$. Thus we assume free or fixed b.c.\ for the lattice.
In two dimensions $d=2$ we studied the correlation function
$G(r)=[(n_i-n_{i+r})^2]_{\rm av}$ for various strengths of the
non-linearity $m$, and found that $G(r)$ increases stronger than
$\log(r)$ with the distance, possibly like $\log^2(r)$, indicating
a superrough low temperature phase, as for $m=2$\cite{sos2}.

To summarize we have presented various applications of a successive
shortest path algorithm for disordered systems containing many
fluxlines or directed polymers with short range repulsion and derived a
number of new results. It would be of high interest to think about
generalization to fluxlines with long range (like Coulomb)
interaction.\cite{taeuber}

This work has been supported by the DFG, and I acknowledge helpful
discussions with M.\ Alava, U.\ Blasum, J.\ Kisker, and
V.\ Pet\"aj\"a.

\end{multicols}

\begin{references}
\bibitem{remark}
  It should be mentioned that the fluxlines
  considered here, although they have hard core repulsion on the bonds,
  can actually touch in discrete lattice points.
\end{references}
\end{document}